\documentclass[mathleft,fleqn,%
]{an}
\usepackage{graphicx}
\usepackage[varg]{txfonts}
\usepackage{color}

\begin{document}

\Pagespan{1}{}
\Yearpublication{2015}%
\Yearsubmission{2015}%
\Month{0}%
\Volume{999}%
\Issue{0}%
\DOI{asna.201400000}%

\title{Towards 21st Century Stellar Models: Star Clusters, Supercomputing
  and Asteroseismology,\thanks{This study uses observational data from
from HST, VLT, AAT, Kepler, and supercomputing resources at iVEC and NCMAS.}}

\author{S.\,W. Campbell\inst{1,2}\fnmsep\thanks{Corresponding author:
        {scampbell@mpa-garching.mpg.de}}
\and T.\,N. Constantino\inst{3,2}
\and V.\,D'Orazi\inst{4}
\and C.\,Meakin\inst{5}
\and D.\,Stello\inst{6}
\and J.\,Christensen-Dalsgaard\inst{7}
\and C.\,Kuehn\inst{8}
\and G.\,M.\,De Silva\inst{9,6}
\and W.\,D.\,Arnett\inst{5}
\and J.\,C.\,Lattanzio\inst{2}
\and B.\,T.\,MacLean\inst{2}
}

\titlerunning{Improving Stellar Models}
\authorrunning{S.\,W. Campbell et al.}

\institute{
Max-Planck-Institut f{\"u}r Astrophysik, Karl-Schwarzschild-Stra{\ss}e 1, 85748
Garching bei M{\"u}nchen, Germany
\and 
Monash Centre for Astrophysics, School of Physics and Astronomy, Monash University, Victoria, 3800, Australia
\and 
School of Physics and Astronomy, University of Exeter, Stocker Road, Exeter EX4 4QL, UK
\and
INAF- Osservatorio Astronomico di Padova, Vicolo dell'Osservatorio 5, 35122
Padova, Italy
\and
Steward Observatory, University of Arizona, 933 N. Cherry Avenue, Tucson AZ 85721, USA
\and
Institute for Astronomy (SIfA), School of Physics, University of Sydney,
NSW 2006, Australia
\and
Stellar Astrophysics Centre, Department of Physics and Astronomy, Aarhus
University, Ny Munkegade 120, DK-8000 Aarhus C, Denmark
\and
Department of Physics and Astronomy, University of Northern Colorado, Greeley, CO, USA
\and
Australian Astronomical Observatory, PO Box 915, North Ryde, NSW 1670, Australia
}

\received{XXXX}
\accepted{XXXX}
\publonline{XXXX}

\keywords{hydrodynamics -- stars: abundances -- stars: evolution -- stars: horizontal-branch -- stars: interiors}

\abstract{Stellar models provide a vital basis for many aspects of
  astronomy and astrophysics. Recent advances in observational astronomy --
  through asteroseismology, precision photometry, high-resolution
  spectroscopy, and large-scale surveys -- are placing stellar models under
  greater quantitative scrutiny than ever. The model limitations are being
  exposed and the next generation of stellar models is needed as soon as
  possible. The current uncertainties in the models propagate to the later
  phases of stellar evolution, hindering our understanding of stellar
  populations and chemical evolution. Here we give a brief overview of the
  evolution, importance, and substantial uncertainties of core helium
  burning stars in particular and then briefly discuss a range of methods,
  both theoretical and observational, that we are using to advance the
  modelling.}

\maketitle

\section{Introduction}
\subsection{On the importance of accurate stellar models}
Stellar models provide various predictions that are widely used in other
fields of astrophysics. For example, they give lifetime predictions for each
phase of evolution (to be compared to star counts), surface properties such
as temperature, luminosity, chemical abundances (to be compared to
photometry and spectroscopy). They also predict contributions to the
interstellar medium (ISM) versus time through stellar mass-loss via winds
and explosions, which release newly formed chemical elements and
electromagnetic radiation. Apart from the external properties and ISM
contributions stellar models also provide internal stellar properties:
thermal structure, chemical profiles, asteroseismic properties, for
example. 
Many of these outputs are key inputs for models of Galactic
evolution and are central to age determinations of stars, both of which are
important for deciphering the processes behind the formation and evolution
of the Milky Way and other galaxies. Any uncertainties in the stellar
models have `knock-on' effects in the understanding of stellar
populations. With the recent advances in observational astronomy -- through
asteroseismology, precision photometry, high-resolution spectroscopy, and
large-scale surveys -- accurate stellar models are under greater
quantitative scrutiny than ever. Here we focus on one of the stages of
evolution that is currently poorly modelled: the core helium-burning (CHeB)
phase.
%
%
\subsection{Overview of core helium burning stars}
Known observationally as red clump, second clump, horizontal branch (HB),
subdwarf B, or RR Lyrae stars -- depending on metallicity, envelope mass,
and total mass -- CHeB stars are numerous and relatively luminous and
therefore contribute disproportionately to Galactic (and extragalactic)
emission. CHeB occurs between the first/red giant branch (RGB) and the
second/asymptotic giant branch (AGB) phase. Lifetimes of CHeB stars are
$\sim 1$ to 10\% of the main sequence lifetimes. Their internal structure
is characterised by dense cores that are (initially) mainly composed of
helium ($\sim 98$\% by mass, left over from core hydrogen burning), a
radiative zone above containing a H-burning shell, and then a convective
envelope (see Fig. \ref{fig:cheb-kipp}). Helium burning occurs in the
convective core and produces C and O. These `ashes' become the C-O cores of
AGB stars and eventually the remnant white dwarves. Subsequent evolutionary
channels of the models, eg. AGB, planetary nebulae, supernovae, is
dependent on the results of this phase, so it is important to simulate this
phase well. Unfortunately, CHeB is where stellar code results start to
diverge significantly, undermining the accuracy of models of supernova
explosions and red giants -- both vital to the chemical evolution of the
Galaxy.

In this paper we first highlight the uncertainties in the current
generation of CHeB models and then briefly outline some of the ongoing
efforts to improve them.
\section{Current state of CHeB modelling}
\begin{figure}
\centering
\includegraphics[width=0.8\linewidth]{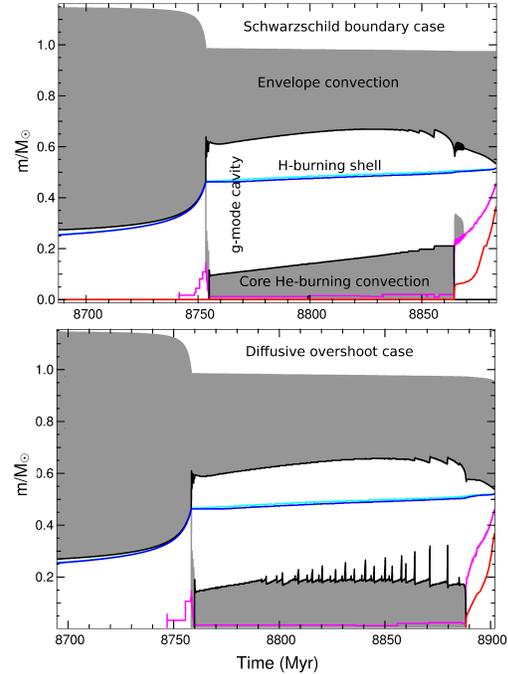}
\caption{Example of the effects of using different convective boundary
  treatments on CHeB evolution in the same stellar evolution code (in
  this case the Monash stellar code MONSTAR; see
  eg. \cite{campbell08}). Both models have an initial mass 1.15
  M$_{\odot}$ and metallicity Z $=0.03$. Apart from the core boundary
  treatment all other inputs are identical. Grey shading represents
  convective regions.  {\bf Top panel:} Schwarzschild boundary with no
  overshoot. We note that there is some core growth in this model. A
  `strict' implementation of the Schwarzschild criterion would result
  in no core growth: see Fig. 2 in \cite{constantino15} for an
  example using the same code. The core growth here is due to the
  numerical treatment at the convective core boundary. This further
  highlights the difficulties of modelling this phase of evolution
  (see also \cite{gabriel14}). {\bf Bottom panel:} Schwarzschild
  boundary with diffusive overshoot. Note the stochastic behaviour of
  the core boundary in the overshoot case.}
\label{fig:cheb-kipp}
\end{figure}

%
%
Stellar codes can give wildly varying evolution results for core
helium burning. In a code comparison between four stellar codes (2007,
private communication) it was found for a 5 M$_{\odot}$ star that (i)
lifetimes differ by up to a factor of two (from $\sim 13$ to 26 Myr),
and (ii) final core masses vary by similar amounts (from $\sim 0.17$
to 0.45 M$_\odot$). The uncertainty holds for massive star models (M
$> 10$ M$_{\odot}$, eg. \cite{langer91}), so it affects (pre)supernova
models and multidimensional simulations that use 1D models as initial
structures. Some of the discrepancies arise from inconsistent
application of the criteria for convective stability (see
\cite{gabriel14} for an analysis) and numerical treatment (eg. top
panel of Fig. \ref{fig:cheb-kipp}). The discordance of the models
points to the fact that we do not yet know how to properly treat
convective boundaries reliably. Moreover CHeB is a particularly
difficult case since an opacity disparity between the convective core
and the radiative zone above builds as helium burning converts He to C
and O. Results from stellar codes are heavily dependent on the
treatment of mixing adopted for the edge of the convective core.  As
an example we show in Figure \ref{fig:cheb-kipp} a pair of our models
that differ only in convective zone boundary treatment (see also
Fig. 15 in \cite{paxton13}). The `spikes' seen in the lower panel
(diffusive overshoot case) are `core breathing pulses' -- sudden
ingestions of helium from the radiative zone due to the numerical
instability of the convective boundary. This mixing reduces the size
of the fully mixed core as compared to the model with no
overshooting. From an asteroseismic point of view this will affect the
oscillation frequencies since the g-mode cavity is different. This is
discussed further below.

%
%
%
\section{Efforts to improve CHeB models}
Our approach to improve CHeB modelling is to use both observations and
theory to inform and constrain the models. Here we give a brief outline of
some of our ongoing work.
\subsection{Star clusters: Spectroscopic observations}
Galactic globular clusters (GC) have long been used to constrain stellar
models. Their colour-magnitude diagrams (CMD) show tight evolutionary
sequences, indicating that they are relatively uniform stellar
populations. Almost all GCs have constant star-to-star abundances of Fe
group elements indicating that the clusters were well-mixed when the stars
formed.  However spectroscopic observations of the light elements (eg. C,
N, O, Na) have revealed that practically all GCs contain at least two
chemically distinct populations (eg. \cite{norris81,gratton12}). The
differences in light-element abundance patterns has been shown to extend
down to the main sequence, indicating that the composition distribution
must have been in place at the earliest times of GC evolution
(eg. \cite{cannon98}).
%
%

The ongoing puzzle of the origins of the GC multiple populations can be
seen as smaller-scale versions of Galactic-scale star formation problems,
especially in chemical space. For example the `Chemical Tagging' technique
(\cite{freeman02}), which quantifies the level of chemical homogeneity
across individual elements in a stellar population, can be used to trace
the origin of Galactic substructure (\cite{desilva06}). Although making GCs
more complex than initially thought, the existence of multiple populations
also opens up other opportunities to study stellar evolution, since the
populations are still much more homogeneous than field star populations. To
exploit this we have made a chemical tagging study of the subpopulations of
NGC 6752 at different evolutionary phases. This showed that stars of
particular composition fail to get to the AGB phase. Instead they appear to
exit to the WD cooling track directly from the CHeB phase, since stars with
high Na are not seen on the AGB. In the case of NGC 6752 70\% of the
cluster stars take this evolutionary path (\cite{campbell13}). Standard
stellar models cannot reproduce this (\cite{campbell13, cassisi14}) and it
represents another shortcoming of CHeB models.
\subsection{Star Clusters: Photometric observations}
\begin{figure}
\centering
\includegraphics[width=0.85\linewidth]{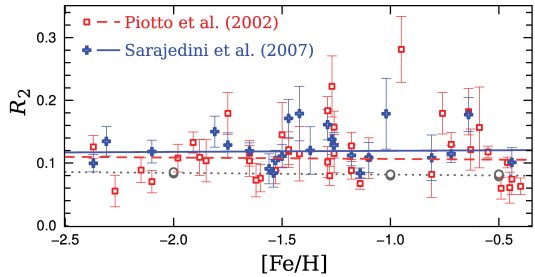}
\caption{Number ratio of AGB stars to CHeB stars (R$_{2}$) versus
  metallicity for 48 GCs. HST photometry is taken from \cite{piotto02} and
  \cite{sarajedini07}, as labelled. The dotted line gives an indication of
  the offset from current models, in this case using the
  semiconvection boundary treatment. See \cite{constantino16} for more
  details}.
\label{fig:R2}
\end{figure}

In Figure \ref{fig:R2} we show our results using star counts of CHeB and
AGB stars in HST photometry of 48 GCs. The number ratio of AGB to HB (CHeB)
stars R$_{2} = N_{AGB}/N_{HB}$ is directly related to the relative
lifetimes of the two phases. We find that R$_{2} = 0.117 \pm 0.005$,
substantially lower than typical previous determinations. Interestingly,
the result is consistent with R$_{2}$ being single-valued across all
GCs. This value is more consistent with some models in the literature,
although it is still somewhat high compared to standard models (see
eg. dotted line in Fig. \ref{fig:R2}).
\begin{figure}
\centering
\includegraphics[width=0.85\linewidth]{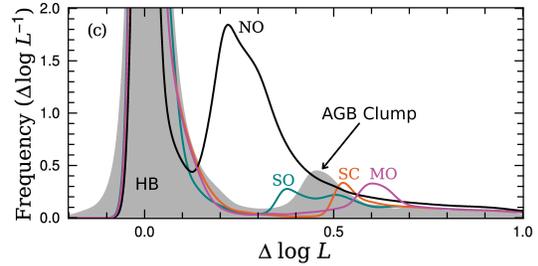}
\caption{Observed average luminosity distribution for 13 GCs that do not
  have an extreme blue extension of their HB (grey shading) and also for
  four stellar models (lines) that differ in convective boundary
  treatment. The models are labelled: NO = no overshoot; SO = standard
  overshoot; SC = semiconvection; MO = maximal overshoot. $\Delta \log L$
  is the luminosity difference to the centre of the HB peak. None of the models
  satisfactorily match the observations. Image adapted from
  \cite{constantino16}.}
\label{fig:LF-EAGB}
\end{figure}

In Figure \ref{fig:LF-EAGB} we show another constraint from photometry that
models must account for, the AGB clump. The observations present a
remarkably sharp peak given that the data is from a host of different GCs
and that there is a range of observational uncertainties, both of which
would tend to widen the peak. Thus this observation is a particularly
strong constraint on models. We also show the effect of using different
boundary treatments during CHeB modelling in the figure. None of the models
satisfactorily reproduce the observation and only the model with no
overshoot can be ruled out definitively. This is another problem for the
CHeB modelling, and highlights the need for multiple constraints. See
\cite{constantino16} for more details on this study.
\subsection{Asteroseismology}
\begin{figure}
\centering
\includegraphics[width=0.8\linewidth]{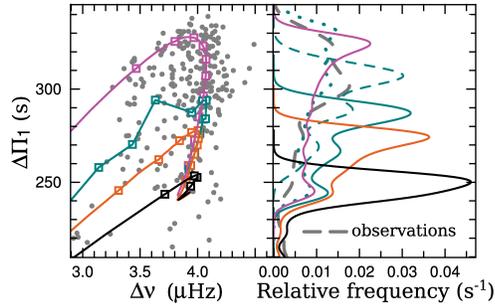}
\caption{Comparison between our suite of models and asteroseismic
  observations of CHeB stars. $\Delta\Pi_1$ is the g-mode period spacing
  and $\Delta\nu$ is the large frequency separation. CHeB models with four
  different convective boundary mixing schemes are shown: no overshoot
  (black), standard overshoot (cyan), semiconvection (orange), and maximal
  overshoot (magenta). Each model has an initial mass 1 M$_{\odot}$. {\bf
    Left panel:} Lines show the evolution of the models, with markers at 10
  Myr intervals. Grey dots show observations (Kepler field stars from
  \cite{mosser14}), and are limited to those with reported mass $0.8 <$
  M/M$_{\odot} < 1.25$. {\bf Right panel:} Probability density curves for
  the models and observations.}
\label{fig:seismo}
\end{figure}

Asteroseismology presents a unique opportunity to provide information
on the interiors of stars. Data from the Kepler and CoRoT space
telescopes have revealed the presence of oscillation modes of mixed g-
and p-mode character in RGB and CHeB stars (eg. \cite{bedding11}). The
g-mode signatures allow inferences on the core structure of stars. We
have investigated the pulsation characteristics of a suite of CHeB
stellar models using a variety of core boundary mixing treatments
(\cite{constantino15};  also see \cite{bossini15}), and
compared these to the mixed-mode period spacings ($\Delta\Pi_1$) for
CHeB stars deduced from Kepler data (\cite{mosser14}). Pulsation
characteristics of CHeB stars depend strongly on core size and
chemical profiles, both of which are affected by the convective
boundary treatment used. We found that standard models (employing the
usual overshoot or semiconvection treatments) cannot match the
mixed-mode period spacings. Only a new mixing routine, ``maximal
overshoot'' (MO) comes close to observations
(\cite{constantino15}). In Figure \ref{fig:seismo} we show a
comparison between the asteroseismic observations and models. Although
we can match the peak of the distribution with the MO model, we
include no physical basis for such a large extension of the convective
core. Also, none of the models can match the distribution at lower
$\Delta\Pi_1$.  Another possible solution to the mismatch between the
models and observations is that the assumptions in
observationally-determined values may be wrong, thereby skewing the
distribution. There is still great potential in using asteroseismology
to constrain the models. It is however very important to know the
observational sample biases in order to make strong conclusions on
stellar structure.
\subsection{Supercomputing: 3D Hydrodynamics models}
\begin{figure}
\centering \includegraphics[width=0.6\linewidth]{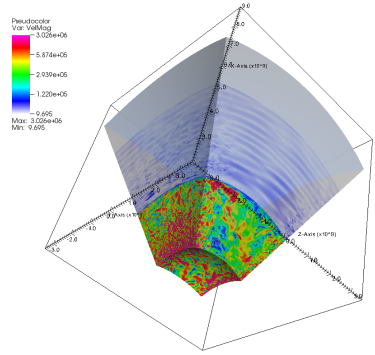} 
\caption{Snapshot of one of our 3D stellar hydrodynamics
  calculations. This simulation models the turbulent convection deep inside
  the helium-burning core of an 8 M$_{\odot}$ star. Colour scale is
  velocity magnitude (cm/s). The regular wave pattern in the stable gas
  above the turbulent core are gravity waves. Spatial axes are in
  centimetres.}
\label{fig:hydro}
\end{figure}

The key problems with CHeB models are related to the treatment of
convection and convective boundaries -- both crudely modelled in 1D at the
moment.  3D hydrodynamics models have the potential to provide physical
insights into these processes (see eg. \cite{meakin07,viallet15}). This new
knowledge could then be used in the 1D models (`321D',
eg. \cite{meakin07,arnett15}), which are still necessary for simulating the
whole lifetime of a star. The 3D models are still hugely computationally
expensive but it is possible to now simulate a few convective turnover
times at low resolution (compared to the resolution needed to resolve all
scales of turbulence). Using the stellar hydrodynamics code PROMPI
(\cite{meakin07}) we have begun a simulation of the CHeB phase in an 8
M$_{\odot}$ star (Fig. \ref{fig:hydro}).
\section{Summary}
It is critical to model stars well since the uncertainties in the stellar
models will affect the interpretation of a range of observations, and feed
back on the modelling and understanding of stellar populations. Given the
current uncertainty of CHeB modelling we advise caution when grids of
stellar models are used in Galactic modelling, particularly for the late
phases of stellar evolution. At least 4 different problems have been
identified for CHeB stars here.  Initial work in improving/constraining the
models is revealing that: (i) it is very difficult to simultaneously match
multiple observational constraints, (ii) biases in observational samples
need to be reported so the data can be used to reliably constrain
models. All the work mentioned in this paper is ongoing. Also of interest
here is the recent analytical work by \cite{spruit15} who estimates the
entrainment rate of He at the convective boundary of CHeB stars, which he
finds is primarily a function of the convective luminosity. This will need
to be included in stellar structure calculations and compared to
observations.

Finally, we are also working on a unique opportunity provided by the new
Kepler space telescope mission K2 -- the seismic observation of a globular
cluster (M4). Here we aim to combine photometry, spectroscopy and
asteroseismology in a well-characterised and (relatively) uniform stellar
population. We now have K2 light curves and high-resolution optical spectra
collected with the HERMES spectrograph (\cite{sheinis15}; AAT) for a
sample of stars at various evolutionary stages.
%
%
%
%
%
%
%

\end{document}